\documentclass[12pt,preprint]{aastex}

\shorttitle{QSO Disk Temperatures \& Continuum Variability}

\shortauthors{Pereyra et al.}

\begin{document}

\title{Characteristic QSO Accretion Disk Temperatures from
       Spectroscopic Continuum Variability}

\author{Nicolas A. Pereyra\altaffilmark{1}, 
        Daniel E. Vanden Berk\altaffilmark{2},
        David A. Turnshek,
        and D. John Hillier}
\affil{University of Pittsburgh, Department of Physics and Astronomy,
       3941 O'Hara ST, Pittsburgh, PA 15260}
\email{pereyra@bruno.phyast.pitt.edu,
       danvb@astro.psu.edu,
       turnshek@quasar.phyast.pitt.edu,
       jdh@galah.phyast.pitt.edu}

\author{Brian C. Wilhite\altaffilmark{3}
        and Richard G. Kron}
\affil{University of Chicago, Department of Astronomy and Astrophysics,
       5640 South Ellis Avenue, Chicago, IL 60637}
\email{wilhite@astro.uiuc.edu,
       rich@oddjob.uchicago.edu}

\author{Donald P. Schneider}
\affil{Pennsylvania State University,
       Department of Astronomy and Astrophyscics,
       504 Davey Laboratory,
       University Park, PA 16802}
\email{dps@astro.psu.edu}

\and

\author{Jonathan Brinkmann}
\affil{Apache Point Observatory, 2001 Apache Point Road,
       Post Office Box 59, Sunspot, New~Mexico 88349-0059}
\email{brinkmann@nmsu.edu}

\altaffiltext{1}{Also at:
                 Universidad Mar\'\i tima del Caribe,
                 Departamento de Ciencias B\'asicas,
                 Catia la Mar, Venezuela}

\altaffiltext{2}{Current address:
                 Pennsylvania State University,
                 Department of Astronomy \& Astrophysics,
                 525 Davey Lab, University Park, PA, 16802}

\altaffiltext{3}{Current address:
                 University of Illinois,
                 Department of Astronomy,
                 1002 W. Green St., Urbana, IL 61801}

\begin{abstract}

Using Sloan Digital Sky Survey (SDSS) quasar spectra taken at multiple
epochs,
we find that the composite flux density differences in the rest frame
wavelength range 1300--6000\AA \  can be fit by a standard thermal
accretion disk model where the accretion rate has changed from one
epoch to the next
(without considering additional continuum emission components).
The fit to the composite residual has two free parameters: a normalizing
constant and the average characteristic temperature $\bar{T}^*$.
In turn  the characteristic temperature is dependent on the ratio
of the mass accretion rate to the square of the black hole mass.
We therefore conclude that most of the UV/optical variability may be due
to processes involving the disk,
and thus that a significant fraction of the UV/optical spectrum may come
directly from the disk.

\end{abstract}


\section{Introduction}
\label{sec_introduction}

It has long been suspected that QSOs are powered by matter accreting
from a disk onto a supermassive black hole
\citep[e.g.,][and references therein]{lin96,ulr97,mir99}.
Also, it is observed that QSOs typically present a ``flattened''
UV/optical spectra or a ``big blue bump''
\citep[e.g.,][]{shi78,mal82,cam84,elv94,cze03}.
In turn,
the big blue bump has generally been interpreted as thermal disk
emission
\citep[e.g.,][]{elv86,lao89,san89,fio95,gu01,sha02}.
The standard model is a geometrically thin disk with local thermal
emission based on conservation of energy and angular momentum
\citep{sha73}.
The integrated standard disk model continuum presents a ``flattened''
spectrum due to the contribution of different disk surface temperatures
at different radii.
Reasonable agreements are found with the standard disk model and the
UV/optical continuum of QSOs
\citep[e.g.,][]{mal83,cze87,wan88,sun89,lao90,kro91,nat98}.

However,
accretion disk continuum fits to the UV/optical spectra of QSOs
typically require additional components such as the extrapolation of the
X-ray continuum into the UV/optical,
the extrapolation of the infrared continuum,
and the inclusion of a ``small blue bump'' component near 3000\AA \ 
\citep[e.g.,][]{mal83,cze87,wan88,sun89,lao90}.
Also, host galaxy light can contaminate the QSO spectrum at optical and
longer wavelengths.

In this work,
we study the residual UV/optical continuum
(continuum difference)
of QSOs when they vary from a ``faint'' phase to a ``bright'' phase,
rather than the continuum itself at either phase.
\citet{wil05} found that the emission lines vary relatively weakly with
respect to continuum variations;
the residual flux density is dominated by continuum changes.
By analyzing a composite residual spectrum we find that the average
UV/optical flux density difference can be accounted for by a standard
accretion disk that has changed its mass accretion rate.
The composite residual is constructed from hundreds of objects observed
by the Sloan Digital Sky Survey
(SDSS)
\citep{wil05}.
This study of the residual flux assumes that the UV/optical variability
is due to only the disk itself,
thus no additional emission components are considered when analyzing the
residual.
The advantage of this approach is that non-variable emission sources
that could contribute to the continuum do not contribute to the
residual.

We are currently studying individual QSO spectra taken from the SDSS.
If the continuum UV/optical variability in individual objects can also
be accounted for by a standard accretion disk at rest frame wavelengths
longer than Ly$\alpha$ 1215{\AA} emission,
then we may have a method to probe the inner disk region by analyzing
the difference between EUV
\footnote{ In this paper EUV refers to the wavelength range
           200\AA-912\AA.}
rest frame residual spectra and the standard
disk model predictions.
The strongest flux density for a disk under typical QSO parameters would
occur in the EUV,
this emission would be produced in the inner disk region.
In turn,
the strongest deviations of a real disk from the standard disk model 
are expected to occur in the inner disk region due to radiative
transfer effects
(including X-ray irradiation)
\citep[e.g.,][]{hub01}
and departures of the gravitational force from the standard Newtonian
$r^{-2}$ law
\citep[e.g.,][]{fra92,kro99}.

In \S\ref{sec_standard} we briefly discuss the emission distribution
from a standard disk and present general relationships.
We present the form in which the average characteristic disk temperature
$\bar{T}^{*}$ is calculated by analyzing the residual spectrum in
\S\ref{sec_determination}.
In \S\ref{sec_application} we apply the discussions of the previous
two sections to a composite spectrum constructed from SDSS data.
Summary and conclusions are presented in \S\ref{sec_sumcon}.

\section{Standard Disk Model}
\label{sec_standard}

\subsection{General Comments}

We assume in this paper that the QSO UV/optical continuum variability is
solely due to accretion disks,
and that the changes in the accretion disk are due to changes in the
mass accretion rate.
We will assume that the radial emission follows that of the standard
disk model.

It is obvious that the standard disk model is an idealization,
just as the assumption of blackbody emission is an idealization for the
continuum of stellar spectra.
However,
we choose to apply the standard model in the continuum variability
analysis for three reasons.
First,
the standard disk model includes much of the physics involved in
accretion disks,
such as gravity from a large central mass determining the average
velocity fields in the disk,
angular momentum transport,
conversion of gravitational energy to radiative emission
(where in the standard model it is assumed to be locally blackbody),
and monotonic increase of temperature with decreasing radius
(except at radii close to the inner disk radius).
Local blackbody emission is a reasonable first approximation that
accounts for the relatively ``flat'' continuum spectrum observed in
systems where accretion disks are inferred.

Second,
the radial emission distribution of the standard model is independent
of the vertical disk structure and the details that generate viscosity
within the disk;
thus,
in the study of continuum variability,
the standard model can be reduced to the dependence on only two basic
physical parameters,
namely:
the central black hole mass
$M_{bh}$
and the disk mass accretion rate
$\dot{M}_{accr}$.

Third,
although more detailed and probably more realistic models that
calculate the disk continuum have been developed since \citet{sha73}
\citep[e.g.,][]
{cze87,wan88,sun89,lao90,ros92,col93,shi94,sto94,dor96,hub00,hub01},
they do not present significant differences with the continuum of the
standard model at wavelengths greater than $\sim$~3600\AA,
and present the strongest differences at wavelengths lower than
$\sim$~1300{\AA}
\citep[see Fig. 13 in][]{hub01}.
Therefore,
since we are analyzing rest wavelength ranges above 1300{\AA},
we do not expect the conclusions of this work to be significantly
affected by applying the standard model rather than a more detailed
one.

More detailed models clearly show,
for example,
that radiative transfer effects
(that must be present)
can generate significant changes in the disk continuum in the EUV.
Therefore,
if the analysis presented here continues to hold up for individual
objects,
then we may be able to study the emission contribution and structure of
the inner disk region by analyzing the deviations in EUV variability
between the standard model and actual observations.

\subsection{Radial Emission Distribution}
\label{sec_red}

In the standard accretion disk model \citep{sha73},
the disk is assumed to be in a steady state and to be azimuthally
symmetric.
Shear stresses transport angular momentum outwards as the
material of the gas spirals inwards.
Conservation of angular momentum leads to the following expression

\begin{equation}
W 2\pi r^2 - \dot{M}_{accr} \omega r^2 = C \,\,\, ({\rm constant})
\;\;\;\; ,
\end{equation}

\noindent
where $r$ is the radius,
$\dot{M}_{accr}$ is the mass accretion rate,
$\omega$ is the corresponding angular velocity,
and $W$ is defined by

\begin{equation}
W \equiv \int\limits_{-z_o}^{z_o} w_{r \phi} \, dz
\;\;\;\; ,
\end{equation}

\noindent
where $z_o$ is the half thickness of the disk and
$w_{r \phi}$ is shear stress between adjacent layers.
With the additional assumption that the shear stresses
are negligible in the inner disk radius

\begin{equation}
\label{equ_stress}
W = {\dot{M}_{accr} \over 2\pi r^2} ( \omega r^2 - \omega_i r_i^2)
\;\;\;\; ,
\end{equation}

\noindent
where $\omega_i$ is the angular velocity at the inner disk
radius $r_i$. 

Taking into account the work done by shear stresses,
and assuming that as the mass accretes inwards the gravitational energy
lost is emitted locally

\begin{equation}
\label{equ_qdisk_gen}
Q = {1 \over 4\pi r} \; {d \over dr}
           \left[
             \dot{M}_{accr} \left(  {\omega^2  r^2 \over 2}
                         - {GM_{bh} \over r}
                       \right)
             - W 2\pi \omega r^2
           \right]
\;\;\;\; ,
\end{equation}

\noindent
where $Q$ is the radiated energy per area of the disk surface,
$G$ is the gravitational constant,
and $M_{bh}$ is the black hole mass.
Assuming that the disk material follows Keplerian orbits,
i.e.,
the mass of the disk is negligible compared to that of the black hole

\begin{equation}
\label{equ_Kepler}
\omega = \left( {G M_{bh} \over r^3} \right)^{1/2}
\;\;\;\; ;
\end{equation}

\noindent
from equations~(\ref{equ_stress}) and (\ref{equ_qdisk_gen}) one finds

\begin{equation}
\label{equ_qdisk}
  Q(r)
=
  {3 \dot M_{accr} G M_{bh} \over 8 \pi r_i^3}
  \left(r_i \over r \right)^3
  \left[1 - \left({r_i \over r} \right)^{1/2} \right]
\;\;\;\; .
\end{equation}

\noindent
The function $Q(r)$ [eq.~(\ref{equ_qdisk})] was originally
derived by \citet{sha73} for binary systems with an accretion disk
rotating about a black hole.

Further,
assuming that the disk is emitting locally as a blackbody,
the radial temperature distribution of the disk will be
given by 

\begin{equation}
\label{equ_tdisk}
  T(r)
=
  \left\{ {3 \dot M_{accr} G M_{bh} \over 8 \pi r_i^3 \sigma_s}
  \right\}^{1/4}
  \left\{
  \left( r_i \over r \right)^3
  \left[1 - \left({r_i \over r} \right)^{1/2} \right] \right\}^{1/4}
\;\;\;\; ,
\end{equation}

\noindent
where $\sigma_s$ is the Stefan-Boltzmann constant.

Defining the characteristic disk temperature $T^*$

\begin{equation}
\label{equ_tstar}
  T^*
\equiv
  \left\{ {3 \dot M_{accr} G M_{bh} \over 8 \pi r_i^3 \sigma_s}
  \right\}^{1/4}
\;\;\;\; ,
\end{equation}

\noindent
the expression of the radial temperature distribution becomes

\begin{equation}
\label{equ_tdisktstar}
  T(r)
=
  T^*
  \left\{
  \left( r_i \over r \right)^3
  \left[1 - \left({r_i \over r} \right)^{1/2} \right] \right\}^{1/4}
\;\;\;\; .
\end{equation}

To further simplify mathematical expressions,
we define the function $t(x)$

\begin{equation}
\label{equ_smallt}
  t(x)
\equiv
  \left\{ x^{-3} \, \left[1 -  x^{-1/2} \right] \right\}^{1/4}
\;\;\;\; ,
\end{equation}

\noindent
thus

\begin{equation}
\label{equ_tdisksmallt}
  T(r)
=
  T^* \, t(r/r_i)
\;\;\;\; .
\end{equation}

\noindent
Since
$r_i$
is the inner disk radius,
$r \ge r_i$.
It follows that

\begin{equation}
  \max(t(r/r_i))
=
  t(7^2 / 6^2)
=
  {6^{3/2} \over 7^{7/4}}
\approx
  0.488
\;\;\;\; .
\end{equation}

\noindent
Therefore the maximum temperature of a standard disk
$T_{max}$ 
is given by

\begin{equation}
  T_{max}
\equiv
  \max(T(r))
=
  {6^{3/2} \over 7^{7/4}} \, T^*
\approx
  0.488 \, T^*
\;\;\;\; ;
\end{equation}

\noindent
that is,
the maximum surface temperature of a standard disk
$T_{max}$
is approximately one-half of the characteristic temperature
$T^*$.

The total disk luminosity
$L_{disk}$
is given by

\begin{equation}
\label{equ_ldiskintr}
  L_{disk}
=
  \int_{r_i}^{r_f} 4 \pi r \, Q(r) \, dr
\;\;\;\; ,
\end{equation}

\noindent
where
$r_f$
is the outer disk radius.
Thus,
substituting equation~(\ref{equ_qdisk}) into
equation~(\ref{equ_ldiskintr}),

\begin{equation}
\label{equ_ldisk}
L_{disk}
=
{\dot M_{accr} G M_{bh} \over 2 r_i}
  \left\{ 1 - {3r_i \over r_f}
    \left[ 1 - {2 \over 3}
      \left({r_i \over r_f} \right)^{1/2}
    \right]
  \right\}
\;\;\;\; .
\end{equation}

Since we are assuming local blackbody emission,
the luminosity density $F_\lambda$ of the disk is given by

\begin{equation}
\label{equ_fnu}
  F_\lambda
=
  \int_{r_i}^{r_f} \pi \, B_\lambda(T(r)) \, 4 \pi r \, dr
\;\;\;\; ,
\end{equation}

\noindent
where $B_\lambda(T)$ is the blackbody intensity at temperature $T$.
That is

\begin{equation}
\label{equ_fnuexplicit}
  F_\lambda
=
  \int\displaylimits_{r_i}^{r_f}
    { \pi \, { {2 h c^2 \over \lambda^5} \over
               \exp \left( h c \over \lambda \, k T^* t(r/r_i) \right)
                - 1 }
      \, 4 \pi r \, dr }
\;\; \;\; ,
\end{equation}

\noindent
where $h$ is Plank's constant,
$c$ is the speed of light,
and $k$ is Boltzmann's constant.
Thus it follows that

\begin{equation}
\label{equ_ldiskintnu}
  L_{disk}
=
  \int_0^\infty F_\lambda \, d\lambda
\;\;\;\; .
\end{equation}

The radial emission distribution of a standard disk
[eqs.~(\ref{equ_qdisk}), (\ref{equ_tstar}),
 (\ref{equ_smallt}), and (\ref{equ_tdisksmallt})]
and its luminosity density
[eqs.~(\ref{equ_fnuexplicit}) and (\ref{equ_smallt})]
are dependent on only four physical parameters:
the black hole mass $M_{bh}$,
the mass accretion rate $\dot{M}_{accr}$,
the innermost radius of the disk $r_i$,
and the outermost radius of the disk $r_f$.

The nine assumptions used to derive the expressions for the radial
emission distribution of a standard disk are:
1) the disk is steady;
2) the disk is azimuthally symmetric;
3) the gravitational field is Newtonian;
4) the mass of the disk is negligible compared to that of the black
   hole;
5) the mass within the disk follows Keplerian orbits;
6) conservation of angular momentum holds as shear stresses cause the
   disk mass to spiral inwards;
7) the shear stress in the disk at the innermost radius is zero;
8) conservation of energy holds as the loss of gravitational energy is
   converted into radiation;
and 9) the disk surface emission is locally blackbody.

\subsection{Additional Assumptions}

We will apply two additional assumptions that reduce the dependence of
the disk luminosity density to only two parameters:
black hole mass
$M_{bh}$
and mass accretion rate
$\dot{M}_{accr}$.

First,
we shall assume that the outer disk radius 
$r_f$
is much larger than the inner disk radius
$r_i$:

\begin{equation}
  r_f
\gg
  r_i
\;\;\;\; .
\end{equation}

This assumption not only eliminates a free parameter,
but it is justified since most of the UV/optical emission 
is coming from the inner disk region [eq.~(\ref{equ_tdisktstar})].
Beyond a certain radius the disk contribution to the UV/optical
continuum is negligible,
even if the disk were to actually extend to infinity.

After making this assumption,
the expression for disk luminosity
[eq.~(\ref{equ_ldisk})]
becomes

\begin{equation}
\label{equ_ldiskinf}
  L_{disk}
=
  {\dot M_{accr} G M_{bh} \over 2 \, r_i}
\;\;\;\; ,
\end{equation}

\noindent
and the luminosity density
[eq.~(\ref{equ_fnuexplicit})]
is now given by

\begin{equation}
\label{equ_fnuexplicitinf}
  F_\lambda
=
  \int\displaylimits_{r_i}^{\infty}
    { \pi \, { {2 h c^2 \over \lambda^5} \over
               \exp \left( h c \over \lambda \, k T^* t(r/r_i) \right)
                - 1 }
      \, 4 \pi r \, dr }
\;\; \;\; .
\end{equation}

Defining

\begin{equation}
\label{equ_s}
  s
\equiv
  {r \over r_i}
\;\;\;\; ,
\end{equation}

\noindent
and rewriting the integral in equation~(\ref{equ_fnuexplicitinf}) in
terms of variable $s$,
the expression for the luminosity density becomes

\begin{equation}
\label{equ_fnuexplicitinfs}
  F_\lambda
=
  r_i^2 \;
  \int\displaylimits_{1}^{\infty}
    { \pi \, { {2 h c^2 \over \lambda^5} \over
               \exp \left( h c \over \lambda \, k T^* t(s) \right)
                - 1 }
      \, 4 \pi s \, ds }
\;\; \;\; .
\end{equation}

\noindent
The shape of the luminosity density
(i.e., the dependence of the luminosity density on wavelength
       within a multiplying constant)
depends only on one physical parameter,
the characteristic temperature $T^*$ [eq.~(\ref{equ_tstar})].

Since a real disk will have a finite outer radius,
the actual shape of the disk luminosity density,
beyond a certain wavelength,
will depend strongly on the exact position of the outer radius;
however,
if the disk is of sufficient size,
the strong dependence on the outer disk radius will occur at infrared
wavelengths and beyond.
While in the UV/optical continuum,
the difference between taking into account the actual outer disk radius
with respect to assuming an infinite disk would be indistinguishable.

Second,
we shall also assume,
as was done by \citet{sha73},
and as is standard in the literature,
that the innermost disk radius is given by the last stable circular
orbit as determined by General Relativity (GR) under a Schwarzschild
metric:

\begin{equation}
\label{equ_rschwarzschild}
  r_i
=
  {6 \, GM_{bh} \over c^2}
\;\;\;\; .
\end{equation}

The expression for disk luminosity [eq.~(\ref{equ_ldiskinf})]
now becomes

\begin{equation}
\label{equ_ldiskinfrs}
  L_{disk}
=
  {1 \over 12} \; \dot{M}_{accr} c^2
\;\;\;\; .
\end{equation}

\noindent
In other words,
if the outer radius extends to infinity,
and the standard model's expression for the inner disk radius
[eq.~(\ref{equ_rschwarzschild})] 
is correct,
the efficiency of a standard disk is
$1 \over 12$.
The expression of $L_{disk}$ now depends on only one physical parameter,
the mass accretion rate $\dot{M}_{accr}$.
Deviations of the inner disk radius from the standard expression
[eq.~(\ref{equ_rschwarzschild})] may generate corrections to
equation~(\ref{equ_ldiskinfrs}).
In particular,
if one were to assume the innermost disk radius
$r_i$
to be the last stable direct circular orbit as calculated in GR under a
Kerr metric
\citep[rather than a Schwarzschild metric, e.g.][]{bar72},
one would find smaller values for $r_i$
\citep[for a discussion on GR corrections under a Kerr metric for the
radial emission distribution of a steady accretion disk see]
[]{nov73,pag74}.

The expression for the characteristic temperature
[eq.~(\ref{equ_tstar})]
now becomes

\begin{equation}
\label{equ_tstarrs}
  T^*
\equiv
  \left\{ {\dot M_{accr} c^6 \over 576 \, \pi \, G^2 M_{bh}^2 \sigma_s}
  \right\}^{1/4}
\;\;\;\; .
\end{equation}

Substituting equation~(\ref{equ_rschwarzschild}) into
equation~(\ref{equ_fnuexplicitinfs}),
the expression for the luminosity density becomes:

\begin{equation}
\label{equ_fnuexplicitinfsrs}
  F_\lambda
=
  \left( 6 G \over c^2 \right)^2 \; M_{bh}^2 \;
  \int\displaylimits_{1}^{\infty}
    { \pi \, { {2 h c^2 \over \lambda^5} \over
               \exp \left( h c \over \lambda \, k T^* t(s) \right)
                - 1 }
      \, 4 \pi s \, ds }
\;\; \;\; .
\end{equation}

Thus,
we find the explicit expression for luminosity density that we apply
here
[eq.~(\ref{equ_fnuexplicitinfsrs}); eq.~(\ref{equ_smallt})].
From equation~(\ref{equ_fnuexplicitinfsrs}),
it can be seen that the disk continuum shape  depends on only one
physical parameter,
the characteristic temperature
$T^*$.

\subsection{Qualitative Comparison with Observations}

\begin{figure}
\epsscale{1.0}
\plotone{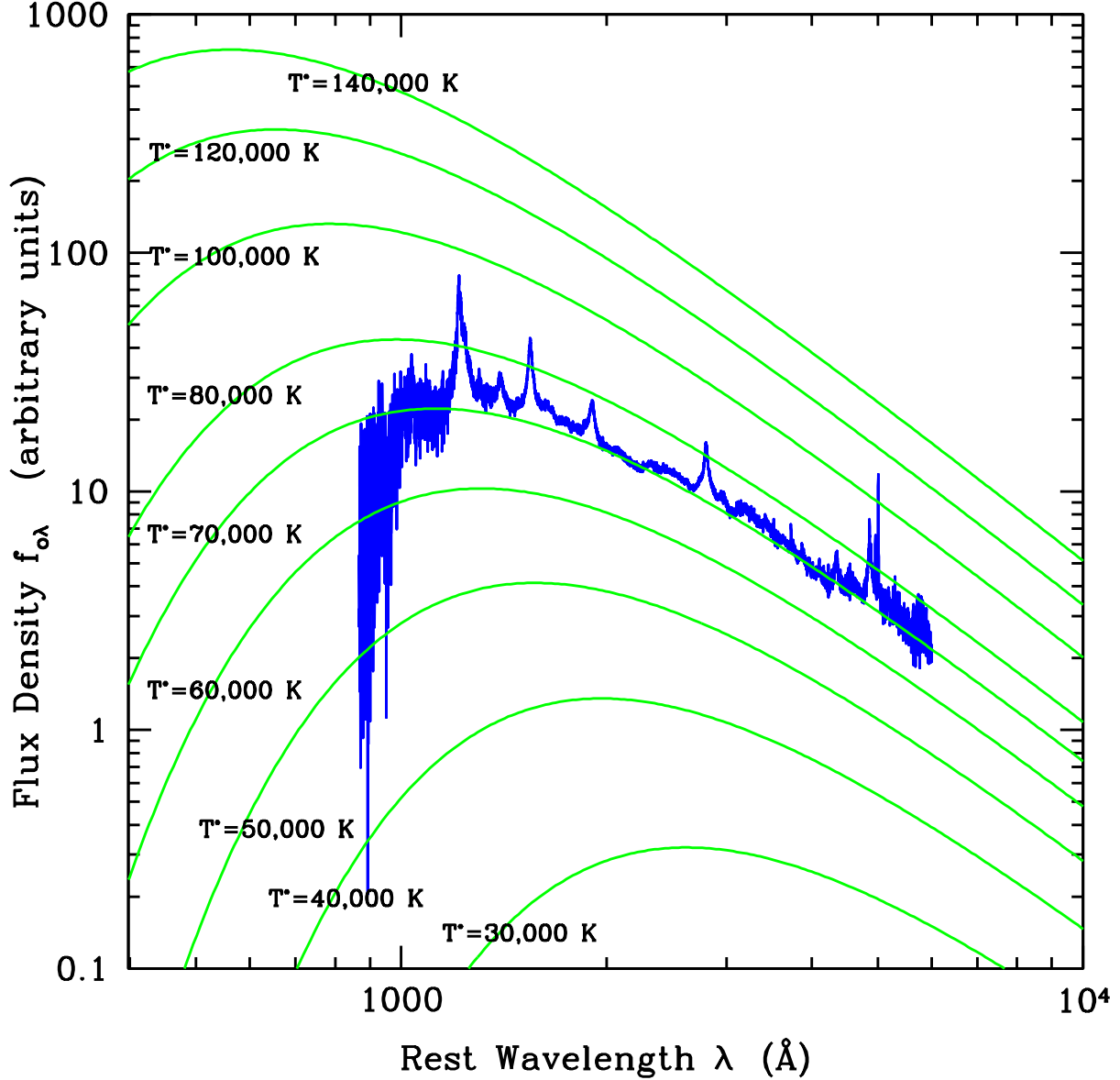}
\caption{
Flux density of an  SDSS QSO composite spectrum
(blue)
in arbitrary units vs. rest wavelength in \AA ,
superimposed with the luminosity density of a standard disk with
different characteristic temperatures $T^*$
(green)
in arbitrary units.
The disk mass accretion rate
$\dot{M}_{accr}$
is proportional to
 $(T^*)^4$
[eq.~(\ref{equ_tstarrs})].
This plot illustrates that that the UV/optical continuum of QSOs is
qualitatively consistent with thermal emission from an accretion disk.
Also,
different $T^*$ values lead to changes in the shape of the disk
continuum spectrum.
}
\label{fig_continuum}
\end{figure}

As discussed in the Introduction,
different authors have noted that the ``flattened'' UV/optical spectra
observed in QSOs is qualitatively consistent with disk models.
To illustrate this point,
we show in Figure~\ref{fig_continuum} the luminosity density calculated
using equation~(\ref{equ_fnuexplicitinfsrs}) for different disk mass
accretion rates superimposed with a QSO composite constructed from SDSS
data
\citep{wil05}.
In Figure~\ref{fig_continuum},
it can be seen that changes in the mass accretion rate
$\dot{M}_{accr}$
produce changes in the form of the disk continuum spectrum.
Therefore,
residual spectra
(i.e.,
 the difference between two spectra observed at different epochs)
can be used to make comparisons with disk models,
as we do here.

The majority of QSOs/AGN present continuum variability on the order of
10\% on timescales of months to years
\citep{sir98,van04}.
These timescales are roughly consistent with estimated sound-speed
timescales of the accretion disks in QSOs/AGN.
Sound-speed timescales measure the time it takes a density/pressure
perturbation
(i.e., a sound wave)
to travel over a significant portion of the disk.
In turn,
the total pressure is equal to the sum of gas pressure and radiation
pressure;
for the standard disk within typical QSO/AGN parameters,
the disk height in the inner disk region is determined by radiation
pressure (rather than gas pressure).
This implies that if one estimates the sound speed solely taking into
account gas pressure
(i.e., taking radiation pressure to be zero),
one significantly underestimates the sound speed in the inner disk
region,
and thus significantly overestimates the sound-speed timescales.

Two additional characteristic disk timescales that should be mentioned
at this point \citep[see e.g.][]{pri81}
are the thermal timescale and the viscous or inflow timescale
\citep[for estimates of different disk timescales for QSO/AGN
       parameters, see e.g.][]{web00}.
The thermal timescale is the ratio of the disk thermal content to its
heating rate for a given patch of disk \citep{kro99}.
The viscous or inflow timescale is the time it takes a disk particle
to travel a significant radial portion of the disk as it spirals
inwards.
The viscous or inflow timescale at a given radius may be estimated by
the ratio of the given radius to the corresponding radial flow
velocity,
and can be thought of as the accretion timescale \citep{kro99}.
Also,
it should be noted that spectroscopic changes occur across the
entire spectrum whenever any part of the disk emission changes.
So,
even if the propagation speed were relatively slow,
since the inner disk dominates the spectrum,
and since the inner disk would change most rapidly,
changes across the spectrum would still occur on relatively short
timescales.

\begin{figure}
\epsscale{1.0}
\plotone{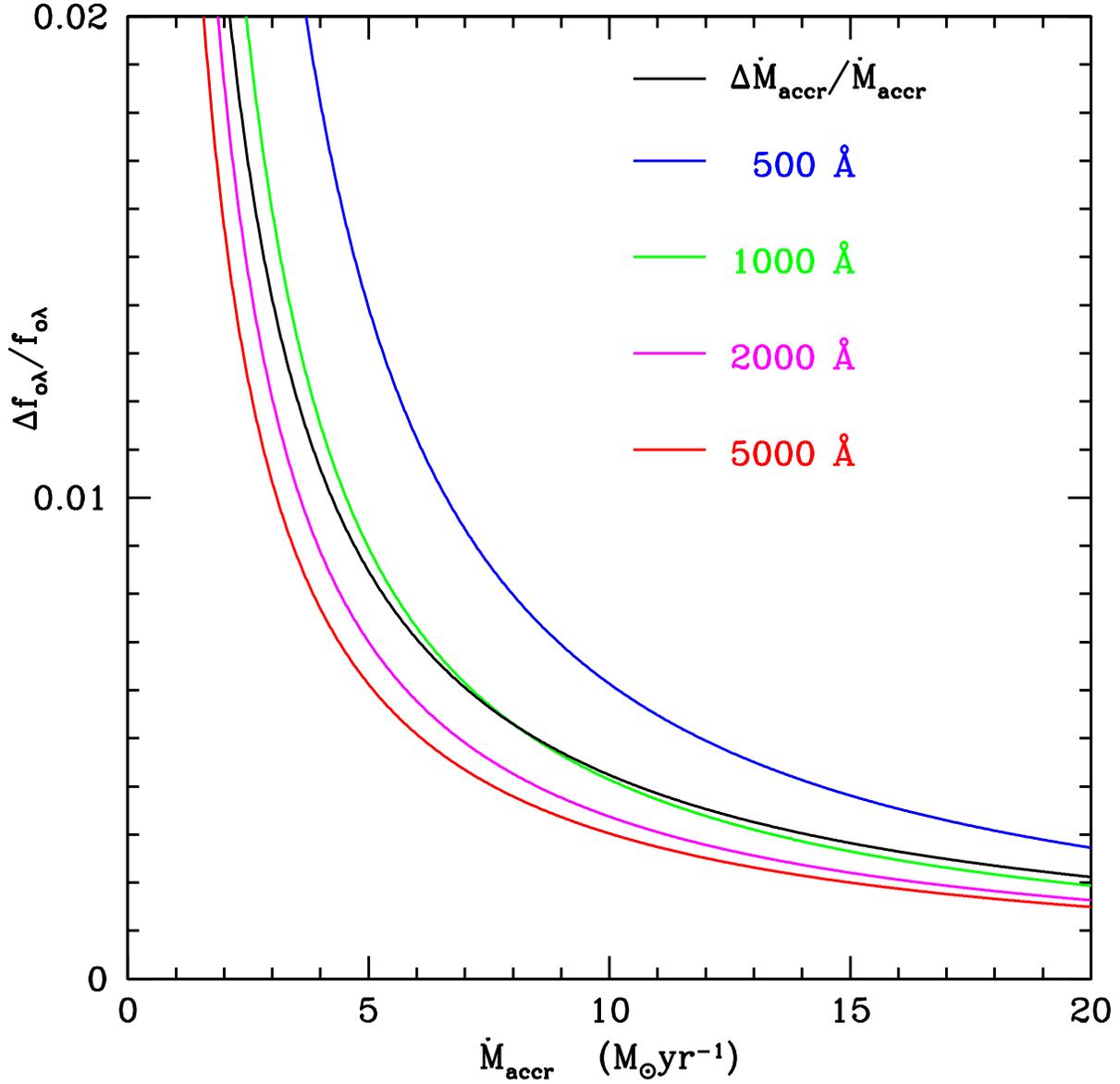}
\caption{
Relative change in flux density of a standard disk vs. mass accretion
rate for different wavelengths. The value of black hole mass used is
$M_{bh} = 10^9 \, M_\sun$.
The change in mass accretion rate is assumed constant
($\Delta \dot{M}_{accr}=0.04 \, M_\sun \, {\rm yr}^{-1}$).
This figure shows that,
within typical QSO parameters,
the continuum spectrum of a standard disk becomes bluer as its mass
accretion rate increases
[or equivalently, as its luminosity increases;
 see eq.~(\ref{equ_ldiskinfrs})].
This is consistent with observations of UV/optical variability
in QSOs,
in that,
as discussed in the text,
the QSO UV/optical continuum typically becomes bluer as it becomes
brighter.
}
\label{fig_color}
\end{figure}

Figure~\ref{fig_color} shows the relative change in flux density of a
standard disk vs. mass accretion rate for different wavelengths.
The value of black hole mass used is
$M_{bh} = 10^9 \, M_\sun$.
The change in mass accretion rate is assumed constant
($\Delta \dot{M}_{accr}=0.04 \, M_\sun \, {\rm yr}^{-1}$).
Figure~\ref{fig_color} shows that,
within typical QSO parameters,
the continuum spectrum of a standard disk becomes bluer as its mass
accretion rate increases
[or equivalently, as its luminosity increases;
 see eq.~(\ref{equ_ldiskinfrs})].
This is qualitatively consistent with observations that show that the
optical/UV continua of QSOs are generally bluer in more luminous phases
\citep[e.g.,][]{cut85,ede90,kin91,pal94,web00,van04,wil05}.

\section{Determination of $\bar{T}^{*}$ from the Residual Spectrum}
\label{sec_determination}

\subsection{General Comments}

In performing fits to the residual spectrum,
we assume that the shape of the disk flux continuum is isotropic
(foreshortening effects will introduce a viewing angle dependence
 to the observed disk flux, but independent of wavelength),
and thus the observed disk flux would be equal to the luminosity density
up to a normalizing constant.
Relativistic effects will introduce dependence of the continuum shape
on viewing angle
(e.g., special relativistic beaming and GR light bending). 
Radiative transfer effects
(e.g., limb darkening)
will also introduce dependence of the continuum shape on viewing angle;
however,
disk atmosphere models suggest that for wavelength ranges above
$\sim$~3600{\AA} significant changes in the continuum shape will occur
only at high viewing angles (close to edge-on).
The strongest changes for wavelengths between
$\sim$~1300 and $\sim$~3600{\AA} will also occur at high viewing angles
\citep[see Fig. 12 in][]{hub00},
and that would include only a fraction of objects assuming a random 
disk orientation in the QSO population.

\subsection{Determination of the Average Characteristic Disk Temperature
            $\bar{T}^*$}
\label{sec_tstardetermination}

To simplify the expressions,
we define

\begin{equation}
\label{equ_gnu}
  g_\lambda(T^*)
\equiv
  \int\displaylimits_{1}^{\infty}
    { \pi \, { {2 h c^2 \over \lambda^5} \over
               \exp \left( h c \over \lambda \, k T^* t(s) \right)
                - 1 }
      \, 4 \pi s \, ds }
\;\; \;\; .
\end{equation}

\noindent
The observed disk flux density $f_{o\lambda}$ will be proportional to
the disk luminosity density
[eq.~(\ref{equ_fnuexplicitinfsrs})].
Therefore,

\begin{equation}
  f_{o\lambda}
=
  c_o \; g_\lambda(T^*)
\;\;\;\; ,
\end{equation}

\noindent
where $c_o$ is a constant that depends on the black hole mass
[see eq.~(\ref{equ_fnuexplicitinfsrs})],
the object's cosmological distance,
and the disk viewing angle
(through foreshortening).

In this work,
we assume that between the two epochs,
the disk evolves from one steady state to another steady state
with a change in mass accretion rate
$\Delta \dot{M}_{accr}$.
Since the black hole mass would not change significantly on the
timescales considered here,
it follows that the normalizing factor
$c_o$
remains constant,
but the characteristic disk temperature $T^*$ will vary
[eq.~(\ref{equ_tstarrs})].
Therefore,
the residual spectrum $\Delta f_{o\lambda}$ is given by

\begin{equation}
\label{equ_deltafonu}
  \Delta f_{o\lambda}
=
  f_{2o\lambda} - f_{1o\lambda}
=
  c_o \; g_\lambda(T^*_2) - c_o \; g_\lambda(T^*_1) 
=
  c_o \; \left( g_\lambda(T^*_2) - g_\lambda(T^*_1)  \right)
\end{equation}

\noindent
where the subindexes $1$ and $2$ correspond to each epoch.

Considering a Taylor series about the average characteristic
temperature

\begin{equation}
  \bar{T}^*
\equiv
  {1 \over 2} \, (T^*_2 + T^*_1)
\;\;\;\; ,
\end{equation}

\noindent
and defining

\begin{equation}
  \Delta T^*
\equiv
  T^*_2 - T^*_1
\;\;\;\; ,
\end{equation}

\noindent
we have

\begin{equation}
  g_\lambda(T^*_2)
=
  g_\lambda(\bar{T}^*)
+ \left( \Delta T^* \over 2 \right)
  \left. \partial g_\lambda \over \partial T^* \right|_{\bar{T}^*}
+ {1 \over 2} \, \left( \Delta T^* \over 2 \right)^2
  \left. \partial^2 g_\lambda \over \partial T^{*2} \right|_{\bar{T}^*}
+ {1 \over 6} \, \left( \Delta T^* \over 2 \right)^3
  \left. \partial^3 g_\lambda \over \partial T^{*3} \right|_{\bar{T}^*}
+ \cdots
\;\;\;\; ,
\end{equation}

\noindent
and

\begin{equation}
  g_\lambda(T^*_1)
=
  g_\lambda(\bar{T}^*)
- \left( \Delta T^* \over 2 \right)
  \left. \partial g_\lambda \over \partial T^* \right|_{\bar{T}^*}
+ {1 \over 2} \, \left( \Delta T^* \over 2 \right)^2
  \left. \partial^2 g_\lambda \over \partial T^{*2} \right|_{\bar{T}^*}
- {1 \over 6} \, \left( \Delta T^* \over 2 \right)^3
  \left. \partial^3 g_\lambda \over \partial T^{*3} \right|_{\bar{T}^*}
+ \cdots
\;\;\;\; .
\end{equation}

\noindent
It follows that

\begin{equation}
  g_\lambda(T^*_2) - g_\lambda(T^*_1)
=
  \Delta T^* \,
  \left. \partial g_\lambda \over \partial T^* \right|_{\bar{T}^*}
+ {1 \over 3} \, \left( \Delta T^* \over 2 \right)^3
  \left. \partial^3 g_\lambda \over \partial T^{*3} \right|_{\bar{T}^*}
+ \cdots
\;\;\;\; .
\end{equation}

\noindent
Therefore,
considering equation~(\ref{equ_deltafonu}),
the residual spectrum
$\Delta f_{o\lambda}$
is given by

\begin{equation}
\label{equ_fnuoexact}
  \Delta f_{o\lambda}
=
  c_o \, \Delta T^* \,
  \left. \partial g_\lambda \over \partial T^* \right|_{\bar{T}^*}
+ {c_o \over 3} \, \left( \Delta T^* \over 2 \right)^3
  \left. \partial^3 g_\lambda \over \partial T^{*3} \right|_{\bar{T}^*}
+ \cdots
\;\;\;\; .
\end{equation}

In this work,
to simplify the analysis,
we introduce the following approximation

\begin{equation}
  \Delta f_{o\lambda}
\approx
  c_o \, \Delta T^* \,
  \left. \partial g_\lambda \over \partial T^* \right|_{\bar{T}^*}
\;\;\;\; .
\end{equation}

The function
$g_\lambda$
[equation~(\ref{equ_gnu})]
depends on only one independent physical parameter,
the characteristic temperature $T^*$.
Therefore,
the process of fitting the residual spectra is reduced to the
determination of two parameters:
the product
$c_o \, \Delta T^*$
and the average characteristic temperature
$\bar{T}^*$.

We determine these two parameters by the standard method of maximum
likelihood under the assumption of Gaussian flux density error
distributions.
That is,
we determine the values of the two free parameters,
``$c_o \, \Delta T^*$''
and
``$\bar{T}^*$,''
that minimize

\begin{equation}
  \chi^2
\equiv
  \sum_{i} { 1 \over \sigma_i^2} \,
            \left(
              \Delta f_{o\lambda i} - 
              [c_o \, \Delta T^*] \,
              \left.
                \partial g_{\lambda i} \over \partial T^*
              \right|_{\bar{T}^*}
            \right)^2
\;\;\;\; ,
\end{equation}

\noindent
where
$\sigma_i$
is the error on the measurement of flux density variation at wavelength
$\lambda_i$.
The standard deviation on the values of each of the two free parameters
is calculated assuming,
once again,
standard Gaussian distributions.

\subsection{Testing the Method Against ``Simulated'' Data}

In order to test the ability of the method described in
\S\ref{sec_tstardetermination}
to determine correct values of $\bar{T}^*$
\footnote{
Assuming of course that a reasonable value for $\chi^2/n$ is found;
where $n$ is the degrees of freedom,
i.e. $n= {\rm \# \ of \ data \ points} -
          {\rm \# \ of \ free \ parameters}$.
},
we shall apply it to ``simulated'' data in which the {\it exact} value
of $\bar{T}^*$ is previously known.

For the test data,
we have fixed the value
$\bar{T}^* = 70,000 \, {\rm K}$
(which corresponds to typical QSO parameters:
 $M_{bh} = 10^9 M_\sun$,
 $\dot{M}_{accr} \approx 1 \; M_\sun \, {\rm yr^{-1}}$),
and have varied the value of $\Delta T^*$ for the assumed constant black
hole mass of
$M_{bh} = 10^9 M_\sun$.
We have also assumed an  error of 5\% on measurement of flux density
at all wavelengths for both the ``bright'' and the ``faint'' epochs.

That is,
we have calculated the luminosity density for
$T^*_2 =  \bar{T}^*  + \Delta T^* / 2$ from
equation~(\ref{equ_fnuexplicitinfsrs})
for
$\bar{T}^* = 70,000 \, {\rm K}$,
for the aforementioned value of black hole mass,
for a discrete number of wavelengths between
1300{\AA} and 6000{\AA},
and taken these to be the measured data points for the bright phase
with a measurement error of 5\% on all points.
Equivalently,
we have also calculated the luminosity density for
$T^*_1 =  \bar{T}^*  - \Delta T^* / 2$ from
equation~(\ref{equ_fnuexplicitinfsrs}) and taken these values to be
the measured data points for the faint phase.
The specific values used for the wavelengths $\lambda_i$ are taken to be
same as those obtained through the construction of the SDSS composite
spectra that is studied in \S\ref{sec_application}.

\begin{figure}
\epsscale{1.0}
\plotone{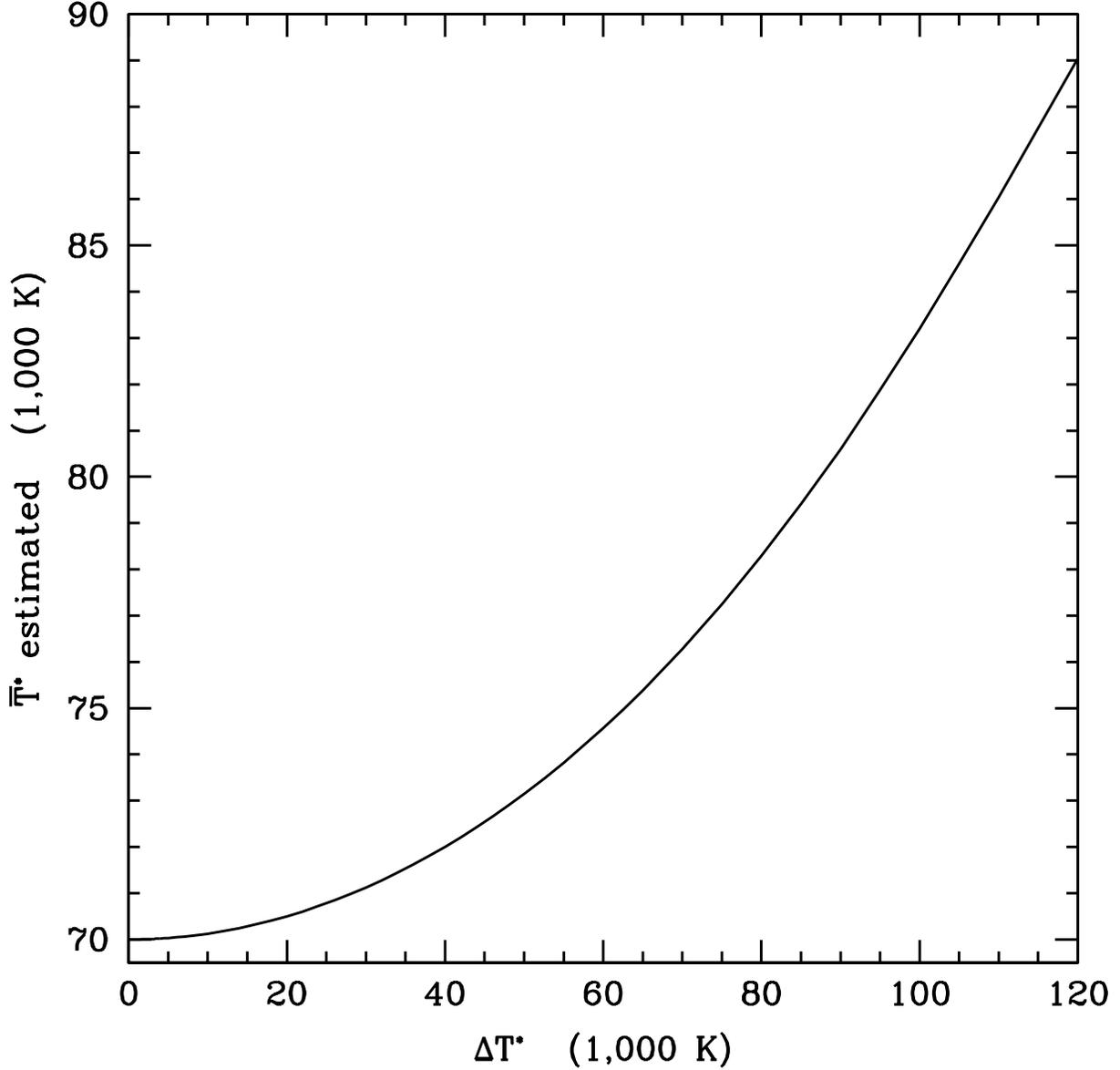}
\caption{
$\bar{T}^*$ estimated through the method discussed in
\S\ref{sec_tstardetermination}
for ``simulated'' data between rest frame wavelengths of
1300{\AA}
and
6000{\AA},
that is,
for simulated data representing the exact residual
flux from the standard disk model
(\S\ref{sec_standard}).
The exact value of
$\bar{T}^*$ is $70,000 \, {\rm K}$,
and a measurement error of 5\% on the simulated density flux
measurements of the bright and faint phase is assumed.
For the above parameters,
accurate values for
$\bar{T}^*$
are obtained for variations in temperature of up to
$\Delta T \approx 15,000 \, {\rm K}$.
That is,
the method estimates accurate values for
$\bar{T}^*$
(for the parameters used here)
for changes in mass accretion rate
$\dot{M}_{accr}$
of up to approximately a  factor of two
[see eq.~(\ref{equ_tstarrs})].
The reason that the characteristic temperature $\bar{T}^*$ becomes
overestimated for larger variations in temperature $\Delta T^*$,
is that the third and higher order terms of
equation~(\ref{equ_fnuoexact})
are no longer negligible.
}
\label{fig_test_ts}
\end{figure}

In Figure~\ref{fig_test_ts},
we show the estimated value
$\bar{T}^*$
vs. the variation in temperature
$\Delta T^*$.
For the above parameters,
accurate values for
$\bar{T}^*$
are obtained for variations in temperature of up to
$\Delta T \approx 15,000 \, {\rm K}$.
That is,
the method estimates accurate values for
$\bar{T}^*$
(for the parameters used here)
for changes in mass accretion rate
$\dot{M}_{accr}$
of up to approximately a  factor of two
[see eq.~(\ref{equ_tstarrs})].
The reason that the characteristic temperature $\bar{T}^*$ becomes
overestimated for larger variations in temperature $\Delta T^*$
(see Figure~\ref{fig_test_ts}),
is that the third and greater order terms of
equation~(\ref{equ_fnuoexact})
are no longer negligible.

\begin{figure}
\epsscale{1.0}
\plotone{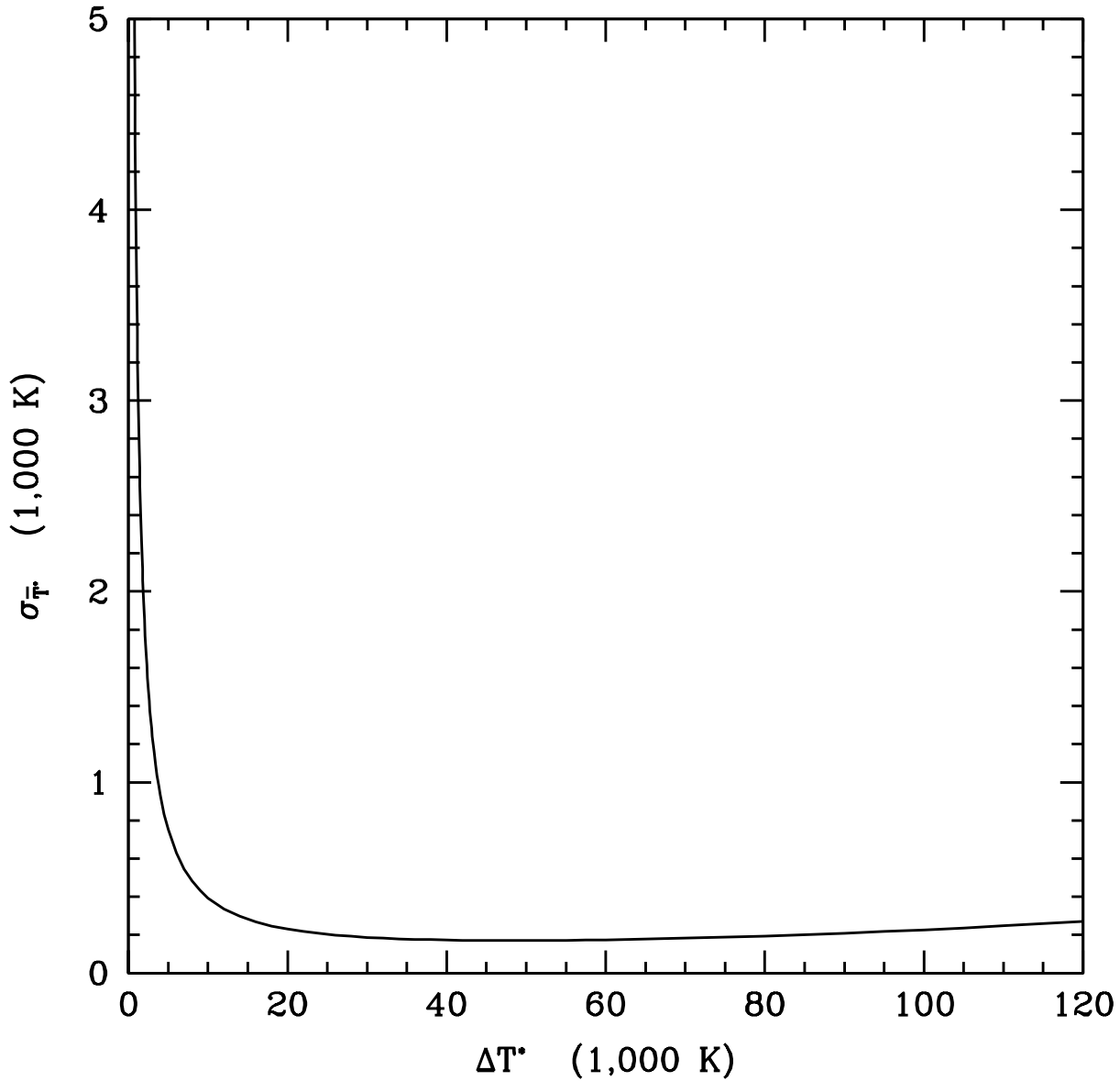}
\caption{
Standard deviation $\sigma_{\bar{T}^*}$ on the estimated value
of $\bar{T}^*$ obtained through the method discussed in
\S\ref{sec_tstardetermination}
for ``simulated'' data as described in and for the same physical
parameters of Figure~\ref{fig_test_ts}.
The exact value of
$\bar{T}^*$ is $70,000 \, {\rm K}$.
For the above physical parameters,
the standard deviation
$\sigma_{\bar{T}^*}$
is less than
$15,000 \, {\rm K}$
for values of
$\Delta T^*$
greater than
$\approx 250 \, {\rm K}$.
The reason that $\sigma_{\bar{T}^*}$ becomes significantly large
(i.e., comparable to the value of
 $\bar{T}^*$)
at very low temperature variations $\Delta T^*$,
is that errors on the flux measurements of the bright and faint phases
become larger than the residual flux.
}
\label{fig_test_sts}
\end{figure}

In Figure~\ref{fig_test_sts},
we show the standard deviation of the estimated value of
$\bar{T}^*$,
$\sigma_{\bar{T}^*}$
vs. the variation in temperature
$\Delta T^*$.
For the above physical parameters,
the standard deviation
$\sigma_{\bar{T}^*}$
is less than
$15,000 \, {\rm K}$
for values of
$\Delta T^*$
greater than
$\approx 250 \, {\rm K}$.
The reason that $\sigma_{\bar{T}^*}$ becomes significantly large
(i.e., comparable to the value of
 $\bar{T}^*$)
at very low temperature variations $\Delta T^*$,
is that errors on the flux measurements of the bright and faint phases
become larger than the residual flux density.

Thus,
we have shown that for a disk characteristic temperature of
$70,000 \, {\rm K}$,
if the UV/optical variability in QSOs is due to changes in mass
accretion rate
$\dot{M}_{accr}$,
then the method described in
\S\ref{sec_tstardetermination}
will be capable of accurately and effectively constraining the average
disk characteristic temperature
$\bar{T}^*$,
as long as the temperature variation
$\Delta T^*$
is greater than
$\approx 250 \, {\rm K}$
and the changes in mass accretion rate
$\dot{M}_{accr}$
are not significantly greater than about a factor of two.

\section{Application to SDSS Composite Spectra}
\label{sec_application}

\subsection{Construction of the Composite Residual Spectrum}
\label{sec_composite}

The systematic construction of the composite residual spectrum from over
300 QSOs taken from the SDSS is discussed in detail in a separate paper
\citep{wil05}.
To avoid redundancy,
we shall present here only a brief discussion of the construction of the
composite,
and focus rather on analyzing the composite residual within the
assumption that it is generated by a change in mass accretion rate of a
standard disk that evolves from one steady state to another,
applying the methods discussed above.

\begin{figure}
\epsscale{1.0}
\plotone{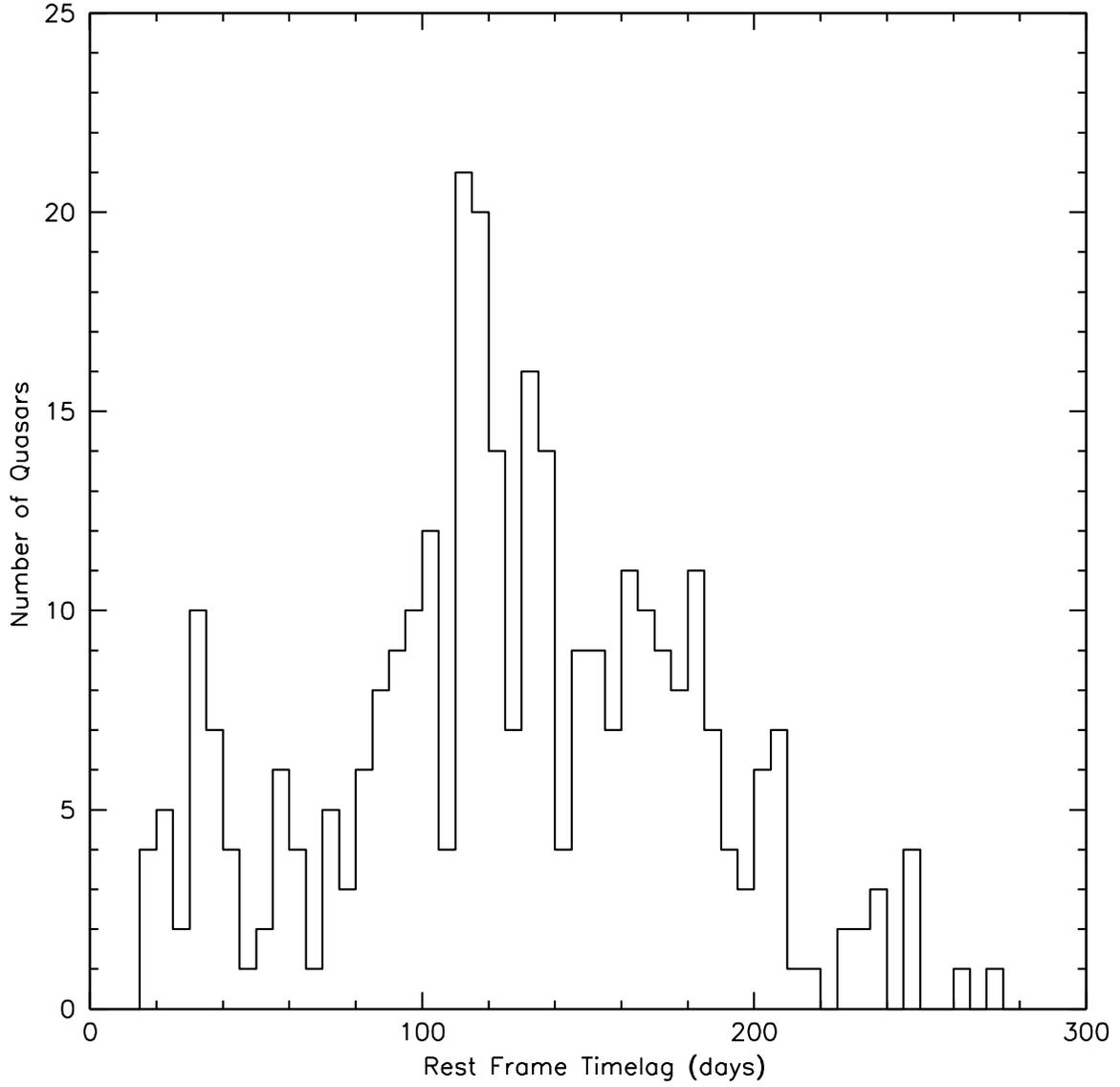}
\caption{
Rest frame time lag histogram of the SDSS QSO object sample used in the
contruction of the composite residual spectrum.
The median rest frame time lag is 124.5 days, and the full sample
ranges from 17.5 to 270.4 days.
}
\label{fig_timelag}
\end{figure}

The SDSS
\citep[see][and references therein for a technical summary]
      {yor00,sto02}
is producing a large homogeneous sample of spectroscopically observed
QSOs
\citep{sch03}.
The QSOs are targeted
\citep{bla03}
according to the algorithm described by
\citet{ric02}
from photometric data in the SDSS imaging survey
\citep{fuk96,gun98,hog01,smi02,pie03,ive04}.
Most of the QSOs used by
\citet{wil05}
are identified in the third SDSS data release
\citep{aba05}.
QSOs that were spectroscopically observed at least twice by the SDSS
(see Figure~\ref{fig_timelag} for a rest frame time lag histogram
 of the sample)
were searched for clear evidence of variability in the observer frame
optical
(UV/optical QSO rest frame).
Stars observed simultaneously with the QSOs were used to calibrate the
observed fluxes
(under the assumption that the stellar fluxes are constant),
and to systematically establish the maximum variation in flux density
measurements that could be attributed to experimental noise.

\begin{figure}
\epsscale{1.0}
\plottwo{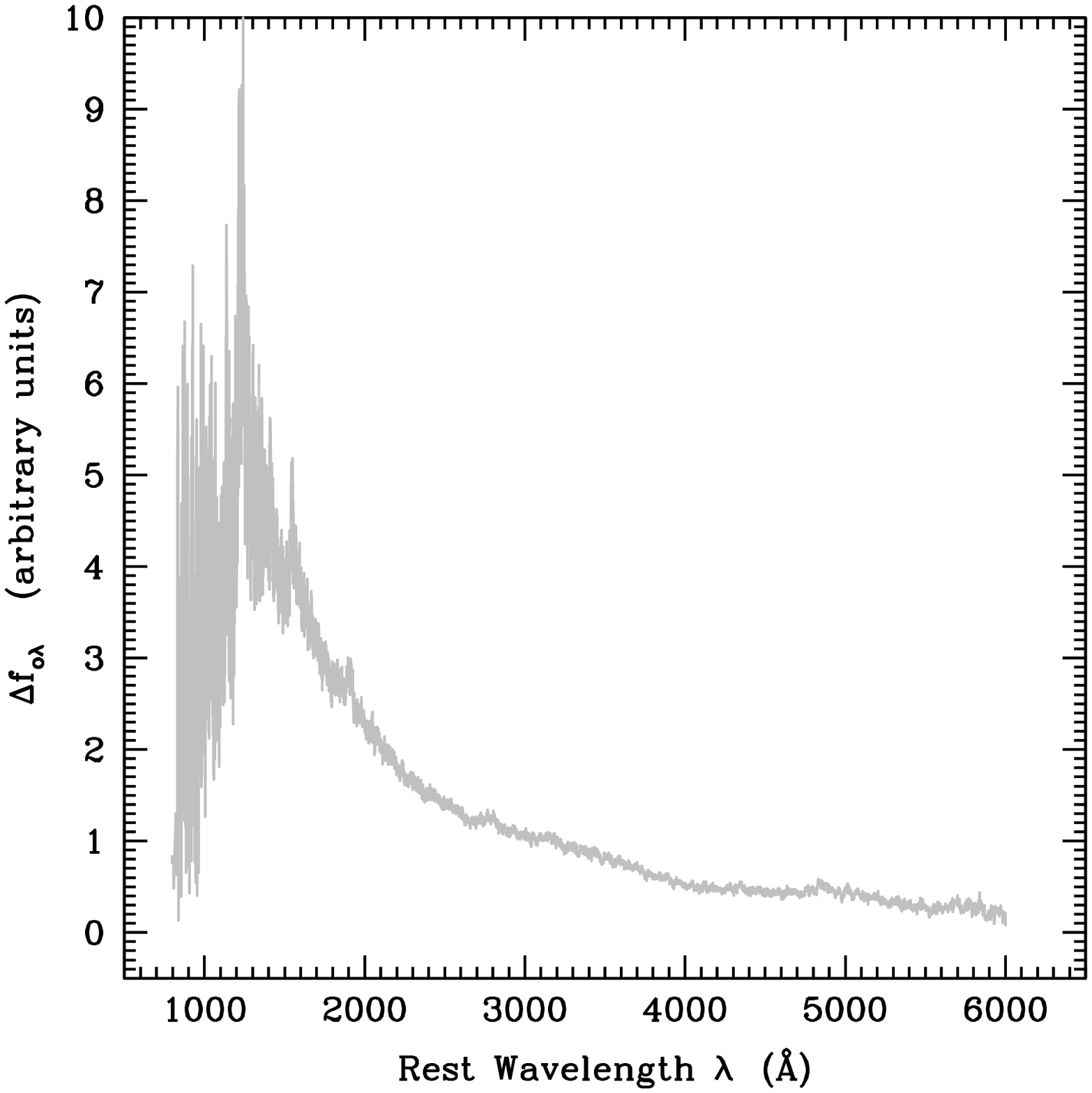}{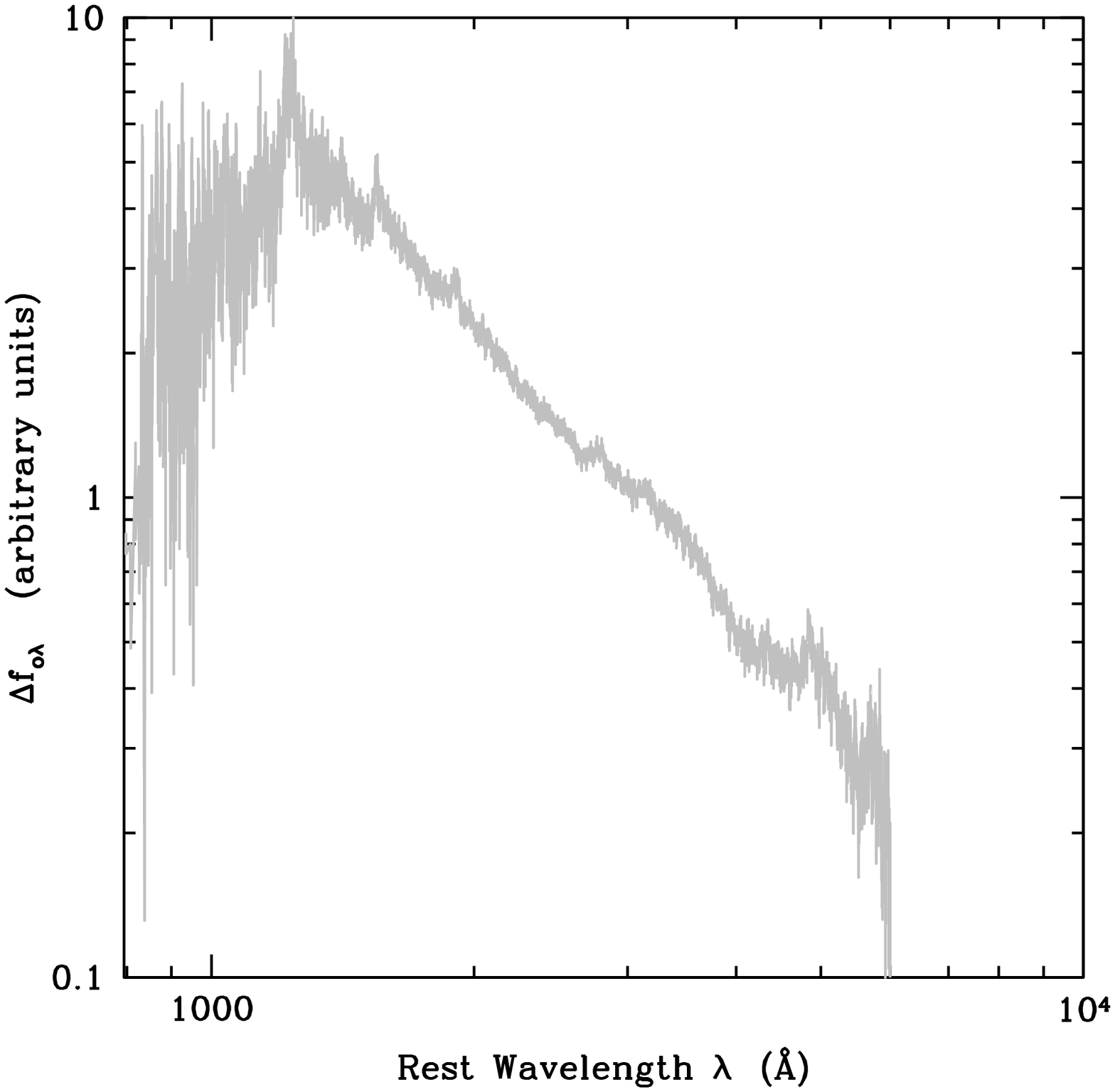}
\caption{
(Left) Composite QSO residual spectrum derived from SDSS data.
(Right) Same plot in log-log scale.
Details on the construction of the composite are presented in a separate
paper \citep{wil05}.
The flux density of the residual spectrum tends to become
larger at lower wavelengths.
Although the strength of the emission lines is significantly reduced in
the residual spectra,
with respect to corresponding line strengths in the spectra of either
the bright or faint phases
(e.g., see Figure~\ref{fig_continuum}),
they are still detectable.
}
\label{fig_residual}
\end{figure}

A residual spectrum was calculated for each of the QSOs that were
identified to have clearly presented a variation in its UV/optical rest
frame continuum.
The individual rest frame residual spectra were scaled to unity at a
predefined wavelength (3060\AA).
The composite residual
(Figure~\ref{fig_residual})
was then constructed by averaging the individually scaled residual
spectra.

\subsection{Fitting the Residual Spectrum}

Following the methods discussed in \S\ref{sec_tstardetermination},
we have fit the SDSS composite residual spectrum discussed in
\S\ref{sec_composite}.
As discussed in \S\ref{sec_tstardetermination},
the two free parameters of the fit are a constant scaling factor and the
average disk characteristic temperature
$\bar{T}^*$.
As discussed by \citet{wil05},
the strength of the emission line components in the residual are
significantly reduced with respect to the line strengths in the spectra
of either the bright  or faint phases;
this point is illustrated by comparing
Figures~\ref{fig_continuum} and \ref{fig_residual}.
That is,
relative to the continuum variations,
the emission lines vary relatively weakly.
However,
residual features at the locations of some major emission lines are
visible in the composite difference spectrum.

We calculate several different fits to the residual spectrum by using
the whole data spectrum (800-6000{\AA}),
by excluding wavelengths below 912{\AA}
(i.e., below the Lyman limit),
by excluding wavelengths below 1300{\AA}
(to exclude flux contributions from the Ly$\alpha$ 1215{\AA}
 line and to avoid the Ly$\alpha$ forest absorption),
and by excluding the contribution of other strong emission lines.
The results are presented in Table~\ref{tab}.
The first column gives the wavelength ranges used for the fits,
the second column lists strong emission lines whose flux contributions
are excluded from the fits
(due to the wavelength ranges of the first column),
the third column gives the derived average characteristic temperatures
$\bar{T}^*$,
the fourth column gives the derived uncertainties
(2$\, \sigma$)
on
$\bar{T}^*$
(calculated under the standard assumption of Gaussian distributions),
and the fifth column gives the reduced chi-square values
$\chi^2 / n$
(where $n$ is the number of degrees of freedom
 [\# of data points - \# of free parameters]).

\begin{figure}
\epsscale{1.0}
\plotone{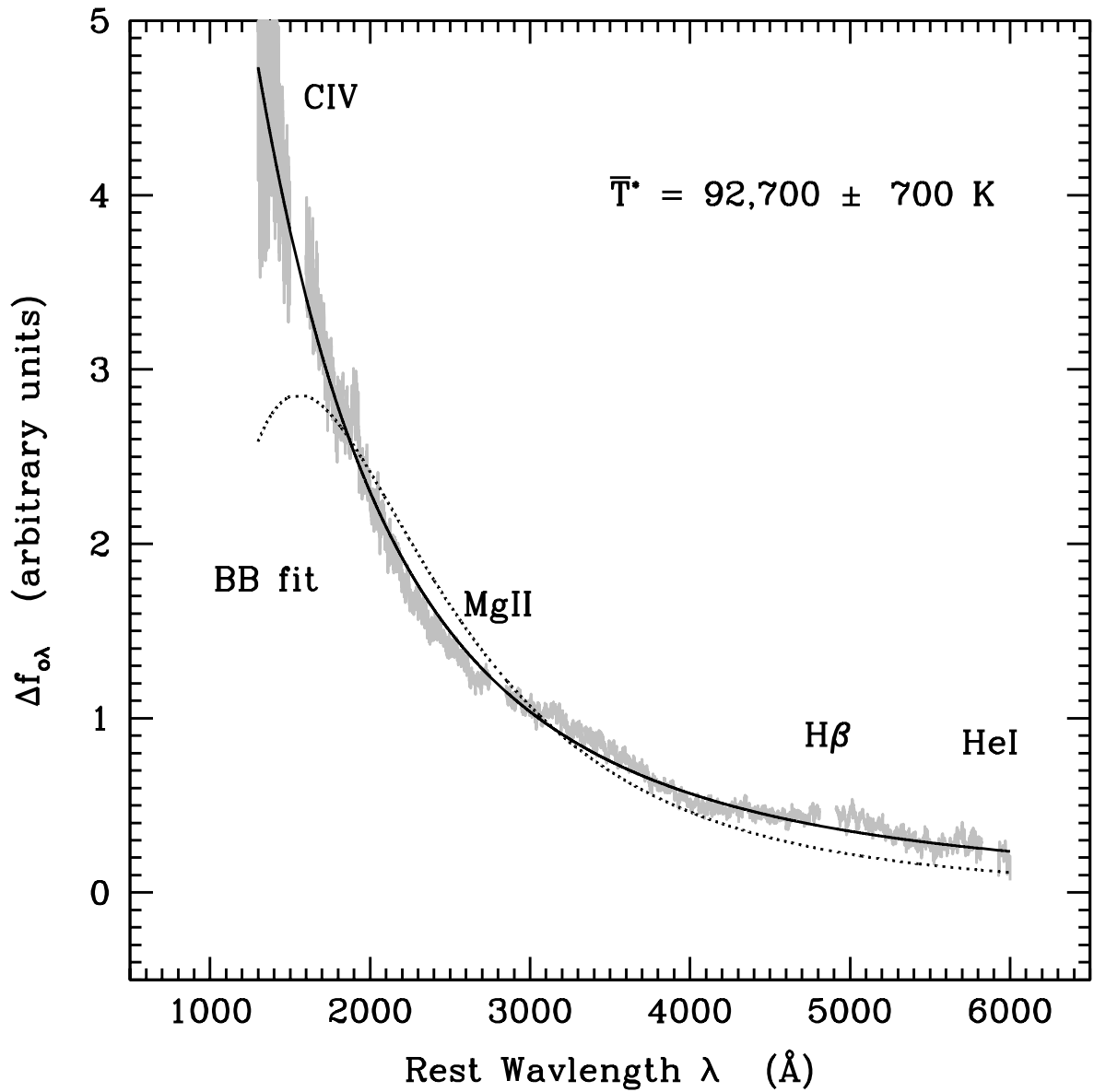}
\caption{
Composite QSO residual spectra derived from SDSS data,
excluding  wavelengths lower than 1300 \AA \  and excluding wavelengths
associated with the four emission lines indicated,
superimposed with the fit of a disk 
({\it solid curve})
that has changed its mass
accretion rate
$\dot{M}_{accr}$.
The average characteristic temperature of the fit is
$\bar{T}^* = 92,700 \, {\rm K}$
(see last row of Table~\ref{tab}).
The maximum disk surface temperature
$T_{max}$
is approximately one-half of the characteristic temperature
$T^*$
(see \S\ref{sec_red}).
In addition, the best blackbody fit
({\it dotted curve};
 $T_{BB} = 15,500 {\rm K}$)
to the composite residual is also shown.
}
\label{fig_fit}.
\end{figure}

In Figure~\ref{fig_fit},
we present the composite QSO residual spectra derived from SDSS data
excluding  wavelengths lower than 1300{\AA},
and excluding wavelengths associated with the four emission lines
indicated in the second column of the last row of Table~\ref{tab},
superimposed with the fit of a standard disk that has changed its mass
accretion rate
$\dot{M}_{accr}$.
The average characteristic temperature of the fit is
$\bar{T}^* = 92,700 \, {\rm K} \pm 700 \, {\rm K}$
and the value for the reduced chi-square of the fit is
$\chi^2/n = 2.993$
(see the last row of Table~\ref{tab}).
Figure~\ref{fig_fit} shows that the composite residual spectrum is
quantatively consistent with the assumption that the residual is caused
by the change of mass accretion rate
$\dot{M}_{accr}$
in a standard disk.

The best fit to the residual spectrum for a single-temperature
blackbody is also shown in Figure~\ref{fig_fit}.
The blackbody fit with a temperature of $15,500$K is clearly quite poor.
Models for which the difference spectrum might reasonably be described
by a single blackbody,
such as individual supernova explosions or localized hot spots,
are not supported by the composite residual spectrum.

Almost all the data used here are different subsets of wavelengths
above the Lyman limit
(912{\AA})
of the QSO rest frame
(Table~\ref{tab}).
Below these wavelengths it is likely that deviations between the
observations and the standard disk model will occur for several reasons.
First,
as discussed in the Introduction,
deviations of a real disk from the standard model are expected below
the Lyman limit due to radiative transfer effects and due to the
departure of the gravitational force from the standard Newtonian form
in the inner disk region.
Second,
for higher redshifts QSOs,
it is more likely that neutral hydrogen intervening systems will produce
significant absorption 
(Ly$\alpha$ forest)
``occulting'' of the disk continuum at wavelengths lower than the
Ly$\alpha$ 1215{\AA}.
A third reason is that other possibly variable components,
such as a Compton upscattering component extending from X-ray
wavelengths,
which do not originate directly from the disk,
will be important.

The average characteristic temperature that we derive from the
composite residual spectra is thus
{\it independent}
of the continuum turnover that may or may not occur
{\it at} $\lambda$ $<$ Ly$\alpha$.
In fact,
most of the fits are based on wavelengths above 1300{\AA}
(Table~\ref{tab}; Figure~\ref{fig_fit}).
The reason this is significant is that the characteristic temperature
of the standard disk model is strongly dependent on the {\it actual}
disk continuum turnover and in particular on the wavelengths where this
turnover occurs.
Thus,
if an observational continuum turnover is introduced at the Lyman
limit due to radiative transfer effects at or near the disk or due to
intervening systems,
rather than due to the radial distribution of disk photospheric
emission,
then attempts to fit the residual including data points below the Lyman
limit would produce inconsistencies that could cause the disk
characteristic temperature to be significantly underestimated.

This analysis has been applied only to a composite residual spectrum,
and so it should give a reasonable value of
$T^*$
for typical QSOs.
However,
as we have not done this for individual QSOs,
we do not know the distribution of
$T^*$,
nor its dependence on luminosity or other QSO parameters.

\begin{figure}
\epsscale{1.0}
\plotone{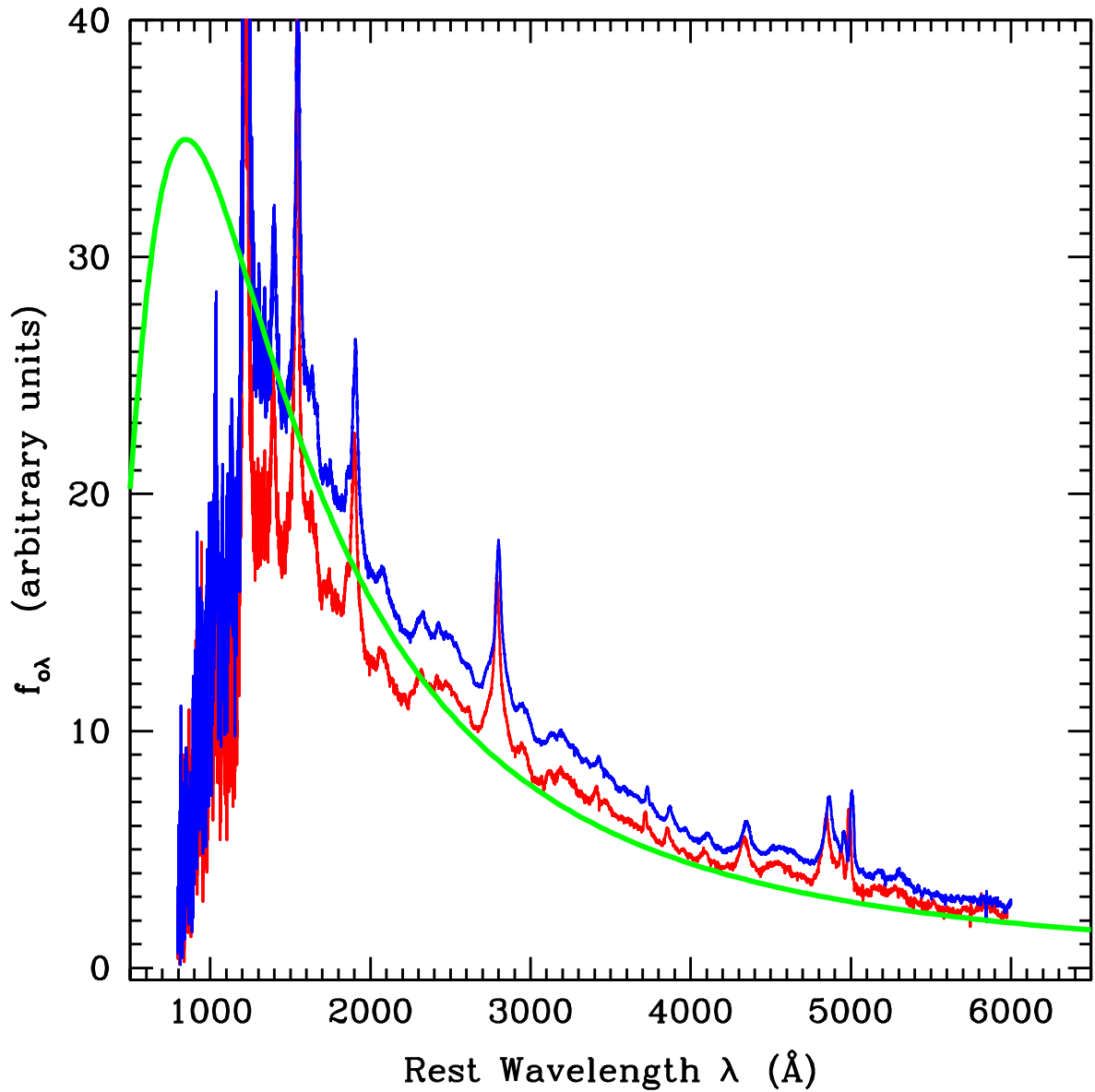}
\caption{
Bright SDSS QSO spectral composite
[blue],
faint SDSS QSO spectral composite
[red],
and the continuum of a standard disk
(up to a normalizing constant)
corresponding to a characteristic temperature of
$T^* = 92,700 \, {\rm K}$
[green]
(see Figure~\ref{fig_fit}).
Although the standard disk model determines an excellent fit to the
composite residual
(Table~\ref{tab},
 Figure~\ref{fig_fit});
when the fit is applied to the composite continuum
(rather than to the residual),
the disk continuum tends to underestimate the observed continuum around
$\sim 3000 \hbox{\rm \AA}$ 
(corresponding to the ``small blue bump''),
and to overestimate at shorter wavelengths.
}
\label{fig_contbf}
\end{figure}

In Figure~\ref{fig_contbf} we show the bright SDSS QSO spectral
composite and the faint SDSS QSO spectral composite \citep{wil05},
superimposed with the continuum of a standard disk
(up to a normalizing constant)
corresponding to a characteristic temperature of
$T^* = 92,700 \, {\rm K}$.
When the accretion disk fit is applied to the composite continuum
(rather than to the residual),
while roughly reproducing the overall UV/optical continuum,
it does not present such an excellent agreement with observations as it
does with the composite residual
(Figure~\ref{fig_fit}).
Within the context of the models that are applied here,
this is due to the contribution of non-variable sources that do not
originate directly from the accretion disk.
For example,
the disk model continuum tends to underestimate the observed emission
around
$\sim 3000 \hbox{\rm \AA}$,
corresponding to the ``small blue bump'' emission,
which is probably composed mainly of Balmer continuum emission and
Fe line complexes.

\section{Summary and Conclusions}
\label{sec_sumcon}

Under the assumption that the UV/optical variability in QSOs is due to
a change of mass accretion rate in a standard disk,
we have analyzed a composite QSO residual spectrum constructed from
hundreds of objects observed by the SDSS.
An advantage of using difference spectra is the ability to remove
non-variable components from the analysis.
We have shown that our analysis technique can recover disk parameters
in simulated data.
The model is relatively simple --- the shape of the difference spectrum
depends only on the disk characteristic temperature
$T^*$.

Our main conclusion is that on average
(i.e., through the SDSS composite),
in the wavelength range 1300-6000{\AA},
residual QSO spectra can be quantitatively accounted for by a standard
thermal disk that has varied from one steady state to another steady
state by changing its mass accretion rate.

In turn,
this suggests that most of the UV/optical variability may be due to
processes involving the disk,
and that a significant fraction of the UV/optical spectrum may come
directly from the disk as has been suggested by many different authors.
 
We are currently studying individual QSO spectra taken from the SDSS.
If the analysis shows consistency with individual objects,
as it has done here for composite spectra,
then we may have a method to probe the inner disk region by analyzing
the difference between variable EUV and the standard disk model
predictions.
Also,
if this method can be used with higher order terms,
then there is a possibility of separating and determining
black hole mass
$M_{bh}$
and disk mass accretion rate
$\dot{M}_{accr}$,
for direct comparison with other methods of measuring these parameters.

\acknowledgments

This work is supported in part by the National Science
Foundation under Grant AST-0071193 and in part by the National
Aeronautics and Space Administration under Grant ATP03-0104-0144.
DVB and DPS were supported by National Science Foundation Grant
AST03-07582.

Funding for the creation and distribution of the SDSS Archive has
been provided by the Alfred P. Sloan Foundation,
the Participating Institutions,
the National Aeronautics and Space Administration,
the National Science Foundation,
the U.S. Department of Energy,
the Japanese Monbukagakusho,
and the Max Planck Society.
The SDSS Web site is http://www.sdss.org/.

The SDSS is managed by the Astrophysical Research Consortium
(ARC)
for the Participating Institutions.
The Participating Institutions are The University of Chicago,
Fermilab,
the Institute for Advanced Study,
the Japan Participation Group,
The Johns Hopkins University,
the Korean Scientist Group,
Los Alamos National Laboratory,
the Max-Planck-Institute for Astronomy (MPIA),
the Max-Planck-Institute for Astrophysics (MPA),
New Mexico State University,
University of Pittsburgh,
University of Portsmouth,
Princeton University,
the United States Naval Observatory,
and the University of Washington.

\clearpage

\begin{deluxetable}{lcccc}

\tablecolumns{5}
\tablewidth{0pt}
\tablecaption{Accretion Disk Fits to the Composite Residual Spectrum
              \label{tab}}
\tablehead{
  \colhead{$\lambda$ (\AA)} \phn &
  \colhead{Excluded Lines} \phn &
  \colhead{$\bar{T}^*$ (K)} \tablenotemark{\dagger} \phn &
  \colhead{2 $\sigma_{\bar{T}^*}$ (K)} \phn &
  \colhead{$\chi^2 / n$} \tablenotemark{\ddagger}}
   
\startdata
\phn 800 - 6000  & \nodata  & 80,100 &   400 & 4.892 \\
\tableline
\phn 912 - 6000  & \nodata  & 82,400 &   400 & 4.695 \\
\tableline
1300 - 6000  & \nodata & 95,500 & 700 & 3.345 \\
\tableline
    1300 - 1500, & \ion{C}{4}  1549{\AA} &
    \nodata & \nodata & \nodata \\
    1600 - 6000  & \nodata     & 91,500   & 700 & 3.111 \\
\tableline
    1300 - 1500, & \ion{C}{4}  1549{\AA} &
    \nodata & \nodata & \nodata \\
    1600 - 2750, & \ion{Mg}{2} 2798{\AA} &
    \nodata & \nodata & \nodata \\
    2850 - 6000  & \nodata    & 91,700   & 700 & 3.155 \\
\tableline
    1300 - 1500, & \ion{C}{4}  1549{\AA} &
    \nodata & \nodata & \nodata \\
    1600 - 2750, & \ion{Mg}{2} 2798{\AA} &
    \nodata & \nodata & \nodata \\
    2850 - 4810, & H$\beta$ 4861{\AA} & 
    \nodata & \nodata & \nodata \\
    4910 - 6000  & \nodata    & 92,700   & 700 & 2.965 \\
\tableline
    1300 - 1500, & \ion{C}{4}  1549{\AA} &
    \nodata & \nodata & \nodata \\
    1600 - 2750, & \ion{Mg}{2} 2798{\AA} &
    \nodata & \nodata & \nodata \\
    2850 - 4810, & H$\beta$ 4861{\AA} & 
    \nodata & \nodata & \nodata \\
    4910 - 5825, & \ion{He}{1} 5875{\AA} &
    \nodata & \nodata & \nodata \\
    5925 - 6000  & \nodata    & 92,700   & 700 & 2.993 \\
			       
 \enddata

\tablenotetext{\dagger}{
The maximum disk surface temperature of a standard disk
$T_{max}$,
is approximately one-half of the characteristic temperature
$T^*$
(see \S\ref{sec_red}).
}

\tablenotetext{\ddagger}{
Reduced chi-square,
where $n$ is the number of degrees of freedom
}

\end{deluxetable}

\end{document}